\documentclass{article}

\usepackage{amsthm,amssymb,amsmath}
\usepackage{graphicx}
\usepackage{epsfig}
\usepackage{cite}

\newcommand{\mh}{M_{h}}
\newcommand{\mH}{M_{H^0}}
\newcommand{\mA}{M_{A^0}}
\newcommand{\mC}{M_{H^\pm}}
\newcommand{\lh}{\lambda_{L}}

\newcommand{\gev}{\mathrm{GeV}}

\newcommand{\mutwo}{\mu_2^2}
\newcommand{\oh}{\Omega h^2}
\newcommand{\ftp}{\Delta_{\oh,\mathrm{total}}}
\newcommand{\gsim}{\lower .75ex \hbox{$\sim$} \llap{\raise .27ex \hbox{$>$}} }
\newcommand{\lsim}{\lower .75ex \hbox{$\sim$} \llap{\raise .27ex \hbox{$<$}} }
\begin{document}
\title{\bf \Large A new viable region of the inert doublet model}

\author{Laura Lopez Honorez\footnote{\tt llopezho@ulb.ac.be}\\
\small Service de Physique Th\'eorique, Universit\'e Libre de Bruxelles, 1050 Brussels, Belgium\\
\and
Carlos E. Yaguna\footnote{\tt carlos.yaguna@uam.es}\\
\small Departamento de F\'{i}sica Te\'orica C-XI and Instituto de F\'{i}sica Te\'orica UAM-CSIC,\\
\small Universidad Aut\'onoma de Madrid, Cantoblanco, E-28049 Madrid, Spain
}

\date{}
%\pacs{}
%\keywords{}
\maketitle
\begin{abstract}
The inert doublet model, a minimal extension of the Standard Model
by a second Higgs doublet, is one of the simplest and most attractive scenarios that can
explain the dark matter.  In this paper, we demonstrate the existence of a new viable region  of the inert doublet model featuring dark matter masses between $M_W$ and about $160~\gev$. Along this previously overlooked region of the parameter space, the correct relic density is obtained thanks to cancellations between different diagrams contributing to dark matter  annihilation into gauge bosons ($W^+W^-$ and $Z^0Z^0$). First, we explain how these cancellations come about and show several examples illustrating the effect of the  parameters of the model on the cancellations themselves and on the predicted relic density. Then, we perform a full scan of the new viable region and analyze it in detail by projecting it  onto several two-dimensional planes. Finally, the prospects for the direct and the indirect detection of inert Higgs dark matter within this new viable region are studied. We find that present direct detection bounds already rule out a fraction of the new parameter space and that future direct detection experiments, such as Xenon100, will easily probe the remaining part in its entirety. 
\end{abstract}
\section{Introduction}
The identification of the dark matter particle is one of the most challenging problems in astroparticle physics today. We know that dark matter exists, for the evidence in its favor is varied and overwhelming. It comes, among others, from rotation curves in galaxies, gravitational lensing effects in clusters of galaxies, the large scale structure of the Universe, and the Cosmic Microwave Background (CMB). The anisotropies in the CMB, in fact, indicate that while dark matter accounts for about $23\%$ of the energy density of the Universe, baryonic  matter reaches only around $4\%$ \cite{Komatsu:2008hk}. Most of the matter in the Universe, therefore, is in the form of dark matter. And yet we still ignore what dark matter consists of.

Currently there is a huge experimental effort aimed at detecting the
dark matter particle and determining its main properties. Dark matter
particles can be detected, for example, via nuclear recoil as they
scatter off  nuclei in \emph{direct} detection experiments.  In recent
years, direct detection experiments such as CDMS~\cite{cdms} or Xenon~\cite{xenon} have been
continuously improving the constraints on the spin-independent dark
matter-nucleon cross section, and they have  started to probe
interesting regions of the parameter space of common dark matter
models. The \emph{indirect} detection of dark matter, which relies on
the annihilation of dark matter particles rather than on their
scatterings, is another promising avenue toward the identification of
the dark matter. Dark matter annihilations may produce photons,
antimatter (positrons and antiprotons mainly) and neutrinos that can
be detected as exotic components in cosmic rays. At present, several
experiments including FERMI (photons)~\cite{fermi}, PAMELA (antimatter)~\cite{pamela} and ICECUBE
(neutrinos)~\cite{ice} are already taking data and setting some constraints on
the dark matter properties. Finally, dark matter particles or other
particles associated with it may also be produced and detected in
colliders such as the LHC. It may be that the LHC soon finds evidence
of physics beyond the Standard Model that sheds some light on the dark
matter puzzle.

In parallel with this experimental effort, there has been a lot of interest  in the theoretical aspects of dark matter. Even though extremely successful in describing collider experiments, the Standard Model of particle physics does not include any viable dark matter candidate. Dark matter, therefore, provides compelling evidence for the existence of new  physics, of physics beyond the Standard Model.  Over the
years, many dark matter candidates have been proposed in different
scenarios for physics beyond the Standard Model. They include the
neutralino~\cite{neutralino} and the gravitino~\cite{gravitino} in supersymmetric
models, the lightest Kaluza-Klein particle in Universal Extra
Dimensional models~\cite{KK}, as well as the  singlet
scalar\cite{singl} or the inert Higgs\cite{Deshpande:1977rw,
  Ma:2006km,Barbieri:2006dq,LopezHonorez:2006gr,Gustafsson:2007pc,Agrawal:2008xz,Lundstrom:2008ai,inertHiggs}
that appear in minimal extensions  of the Standard Model. Thanks to
the simplicity of the underlying model and to the rich phenomenology
associated with it, the inert Higgs --the lightest odd particle of the
inert doublet model-- has earned a special place among them as one of
the most attractive  dark matter candidates considered in the recent
literature.

In the inert doublet model, a Higgs doublet $H_2$, odd under a new
$Z_2$ symmetry, is added to the Standard Model particle content.  The
lightest inert (odd) particle, $H^0$, turns out to be  stable and
hence a suitable dark matter candidate.  After
being introduced in \cite{Deshpande:1977rw}, this model has been extensively studied in a number of recent works \cite{Ma:2006km,Barbieri:2006dq, LopezHonorez:2006gr, Gustafsson:2007pc, Agrawal:2008xz, Lundstrom:2008ai, inertHiggs}. In \cite{Ma:2006km}, it was extended with odd right-handed neutrinos so as
to explain neutrino masses. Then,
in \cite{Barbieri:2006dq}, it was proposed as a possible way to increase the Higgs
mass without contradicting electroweak precision data. Regarding dark matter,  the main features
of the model were systematically analyzed in
\cite{LopezHonorez:2006gr}. Later on, in \cite{Gustafsson:2007pc}, it
was shown that the annihilation of inert Higgs dark matter can give rise
to significant gamma ray lines. Recently, it was demonstrated that, for masses between $60~\gev$ and $M_W$, dark matter annihilations are usually dominated by three-body final states consisting of a real and a virtual $W$ \cite{Yaguna:2010hn,Honorez:2010re}, rather than by the two-body final states considered in previous works. A result that  has important implications for the determination of the viable parameter space of the model and for the detection prospects of inert Higgs dark matter.

These  various studies have established that in the inert doublet
model the dark matter constraint can only be satisfied within
specific regions of the parameter space. Three viable regions have
been identified: a low mass regime, with $\mH\lesssim 50~\gev$, where dark
matter annihilates exclusively into fermion pairs; an intermediate mass regime, with $\mH\lesssim M_W$, where dark matter annihilations are dominated by the final states $WW^*$ and $b\bar b$, and where coannihilations with $A^0$ may also play a role in the determination of the relic density; and a heavy mass regime, $\mH\gtrsim 500~\gev$, where the annihilation into gauge bosons is dominant and the mass splitting between the inert particles is small.

We demonstrate in this paper the existence of another viable region of the inert doublet model, a region  that had been overlooked in previous analysis. This region features dark matter masses between $M_W$ and about $160~\gev$ and is compatible  with  the relic density constraint  thanks to cancellations among the different diagrams that contribute to dark matter annihilation into gauge bosons.  It is our goal to introduce this new viable region and to analyze its main properties. We determine its position in the multidimensional parameter space of the inert doublet model and study its implications for the direct and indirect detection of inert Higgs dark matter. Remarkably, we find that this new viable region will be entirely and easily probed by the next generation of direct dark matter detection experiments such as Xenon100.

In the next section, the inert doublet model is briefly reviewed, with particular emphasis on the different bounds that must be taken into account. Then, we describe in some detail the known viable regions of the model, those that are compatible with the observed dark matter abundance and with all other experimental and theoretical constraints. In section  \ref{sec:canc}, we illustrate, analytically and numerically, the existence of cancellations in the annihilation of inert Higgs dark matter into $W^+W^-$ and $Z^0Z^0$ that allow to satisfy the relic density constraint for dark matter masses above $M_W$. The effect of the different parameters of the model on these cancellations and on the predicted relic density are studied,  and several examples of models within the new viable region are provided. In section \ref{sec:new}, we  use  Markov Chain Monte Carlo techniques to scan the parameter space of the inert doublet model and obtain a large sample of models within the new viable region.  We then determine the main features of this region by projecting this multidimensional  cloud of viable models onto several two-dimensional planes.  The issue of the fine-tuning required to obtain the correct relic density is also addressed, and it is found that these models are not particularly fine-tuned. Finally, we consider, in section  \ref{sec:det}, the detection prospects of inert Higgs dark matter in direct and indirect detection experiments.  
\section{The inert doublet model}
The inert doublet model is a simple extension of the Standard Model
with one additional Higgs doublet $H_2$ and an unbroken $Z_2$ symmetry
under which $H_2$ is odd while all other fields are even. This
discrete symmetry prevents the direct coupling of $H_2$ to fermions and,
crucial for dark matter, guarantees the stability of the lightest
inert particle. The scalar potential of this model is given by
\begin{align}
\label{eq:V}
V=&\mu_1^2|H_1|^2+\mu_2^2|H_2^2|+\lambda_1|H_1|^4+\lambda_2|H_2|^4+\lambda_3|H_1|^2|H_2|^2\nonumber\\
&+\lambda_4|H_1^\dagger H_2|^2+\frac{\lambda_5}{2}\left[(H_1^\dagger H_2)^2+\mathrm{h.c.}\right]\,,
\end{align}
where $H_1$ is the Standard Model Higgs doublet, and $\lambda_i$ and
$\mu_i^2$ are real parameters. As in the Standard Model, $\lambda_1$ and $\mu_1^2$ can be traded for the Higgs mass, $\mh$, and the vacuum expectation value of $H_1$, $v=246~\gev$\footnote{Notice that in \cite{Barbieri:2006dq} a different  convention, $v=174~\gev$, is used.}. Four new physical states are obtained in
this model: two charged states, $H^\pm$, and two neutral ones, $H^0$
and $A^0$. Either of them could be dark matter. In the
following we choose $H^0$ as the lightest inert particle,
$\mH^2<\mA^2,\mC^2$, and, consequently, as the dark
matter candidate. After electroweak symmetry breaking, the inert scalar
masses take  the following form
\begin{align}
\label{eq:masses}
\mC^2&= \mu_2^2+\frac12\lambda_3v^2 , \\
\mH^2&= \mu_2^2+\frac12(\lambda_3+\lambda_4+\lambda_5)v^2 , \\
\mA^2&= \mu_2^2+\frac12(\lambda_3+\lambda_4-\lambda_5)v^2\,. 
\end{align}
Notice that $\lambda_5$ determines the mass splitting between $\mH$ and $\mA$ and is necessarily negative as $\mA>\mH$. The mass splitting between the charged and the neutral components, on the other hand, is due to both  $\lambda_4$ and $\lambda_5$. It is convenient to trade for $\mutwo$ the parameter $\lh$ defined by 
\begin{equation}
\label{eq:lh}
\lh\equiv\frac{\lambda_3+\lambda_4+\lambda_5}{2}=\frac{\mH^2-\mutwo}{v^2}\,.
\end{equation}
As we will see, this parameter, which  determines the interaction term between  a pair of $H^0$ and the Higgs boson, will play a crucial role in the cancellations discussed in Section \ref{sec:canc}.  The free parameters of the inert doublet model can then be taken to be the following:
\begin{equation}
\mH,\mC,\mA,\mh, ~\mathrm{and}~\lh\,.
\end{equation}
From them the $\lambda_i$ and $\mu_i^2$ parameters appearing in the scalar potential, equation (\ref{eq:V}), can be reconstructed via equations (\ref{eq:masses}-\ref{eq:lh}).

These parameters  are subject to a number of theoretical and experimental constraints --see  \cite{Barbieri:2006dq} and
\cite{LopezHonorez:2006gr}. To prevent the breakdown of perturbation theory, the $\lambda_i$ parameters  cannot be very large.  We demand accordingly that
\begin{equation}
|\lambda_i|<4\pi\,.
\end{equation}
The requirement of vacuum stability, in addition, imposes the following conditions on the $\lambda_i$ parameters:
\begin{equation}
\lambda_1,\lambda_2 > 0\,,\qquad 
\lambda_3, \lambda_3+\lambda_4-|\lambda_5|>-2\sqrt{\lambda_1\lambda_2}\,.
\label{eq:stab}
\end{equation}
Since $\lambda_2$ only affects the quartic interaction term among the inert particles, the phenomenology of the model is rather insensitive to its precise value.  For definiteness we set $\lambda_2=3.0$ throughout this work and use the above equation as a constraint on the remaining parameters of the model.

Accelerator bounds should also be taken into account. The mass of the charged scalar, $\mC$, is constrained  to be larger than $70-90\,\gev$ \cite{Pierce:2007ut}; $\mH+\mA$ must exceed $M_Z$ to be compatible with the $Z^0$-width measurements; and  some regions in the plane ($\mH,\mA$) are constrained by LEP II data \cite{Lundstrom:2008ai}.  Another important restriction on the parameter space of the inert doublet model comes from electroweak precision data. The inert doublet, $H_2$,  contributes to electroweak precision parameters such as $S$ and $T$. It turns out that its contribution  to $S$ is always small \cite{Barbieri:2006dq}, so it can be neglected. Its contribution to $T$, $\Delta T$, was computed in \cite{Barbieri:2006dq} and is given by
\begin{equation}
\Delta T=\frac{1}{16\pi^2\alpha v^2}\left[F(\mC,\mA)+F(\mC,\mH)-F(\mA,\mH)\right]
\label{deltat}
\end{equation}
where
\begin{equation}
F(m_1,m_2)=\frac{m_1^2+m_2^2}{2}-\frac{m_1^2m_2^2}{m_1^2-m_2^2}\ln\frac{m_1^2}{m_2^2}\,.
\end{equation}
This contribution originates in the $\lambda_{4,5}$ terms in the potential, for those are the terms breaking the custodial symmetry. The Higgs mass, $\mh$, also affects $T$ via \cite{Peskin:1991sw}
\begin{equation}
T_h\approx -\frac{3}{8\pi\cos^2\theta_W}\ln\frac{\mh}{M_Z}.
\end{equation}  
For large $\mh$ this contribution violates current experimental constraints. In fact, if no new physics affects the electroweak precision parameters, $\mh$ should be smaller than $186~\gev$ at $95\%$C.L. In the inert doublet model, however, there is certainly  physics beyond the Standard Model, so it is possible to have a heavier Higgs boson provided that the inert particles give rise to a positive compensating $\Delta T$. In fact, the Higgs boson can be as heavy as $600~\gev$ in the inert doublet model \cite{Barbieri:2006dq}. 

It is useful to differentiate a light and a heavy Higgs regime in this model. In the heavy Higgs regime, $\mh>200~\gev$, $\Delta T$ should be positive and  equation (\ref{deltat}) implies that  $\mC>\mA$. In the light Higgs regime ($\mh<200~\gev$), on the contrary, there is no correlation between $\mC$ and $\mA$. 

In our numerical analysis we implement the electroweak precision constraint by requiring that 
\begin{equation}
-0.1<\Delta T+T_h<0.2\,,
\label{eq:ewpd}
\end{equation}
in agreement with present bounds \cite{EWPD}.

Finally, since dark matter is the most compelling motivation of the inert doublet model, we also constrain the relic density of inert Higgs dark matter to be compatible with the observed dark matter density.  Explicitly we impose that
\begin{equation}
0.09<\oh <0.13,
\end{equation}
which is consistent with the WMAP measurement \cite{Komatsu:2008hk}. To ensure that the relic density of inert Higgs dark matter is accurately computed, we make use of the micrOMEGAs package \cite{Belanger:2006is} throughout this work.

As expected,  all these constraints can be satisfied only within
certain regions of the parameter space, the so-called \emph{viable}
regions. Before describing the new viable region of the inert doublet
model --the main topic of this paper--, we will have a look in the next section to the previously known viable regions. 

%
%
%
%
%
%
%
%L
\section{The known viable regions}
In previous works \cite{Barbieri:2006dq,Cirelli:2005uq, LopezHonorez:2006gr,Hambye:2007vf}, it has been found that the dark matter
constraint can only be satisfied for restricted values of
$\mH$. Three viable regions can be distinguished: a small mass
regime, with $\mH \lesssim 50~\gev$; a large mass regime, where $\mH > 500$
GeV; and an intermediate mass regime where $50~\gev\gtrsim  \mH<M_W$. 

In the small mass range,  $A^0 $ and $H^{\pm}$ must be decoupled from $H^0$   
in order to satisfy collider and precision test constraints.  Thus, the inert doublet model becomes equivalent (for dark matter purposes) to the singlet
scalar model of dark matter \cite{singl}. Dark matter annihilations proceed in this case through Higgs mediated
 diagrams into light fermions and are therefore dominated by the
 $b\bar b$ final state.  Interesting  phenomenological issues related to this mass range, including prospects for the direct and the indirect detection of dark matter,  were studied in~\cite{Hambye:2007vf,Andreas:2008xy,Andreas:2009hj,Arina:2009um,Nezri:2009jd}. Recently, this small mass regime of the inert doublet model, in particular the region $\mH\sim 8~\gev$,  received some  attention  given the results of DAMA~\cite{Bernabei:2008yi} and Cogent~\cite{Aalseth:2010vx} experiments.

On the other edge of the mass range, the high mass regime, between $535~\gev$ and $50$ TeV,  is quite generic for scalar candidates of  minimal dark matter
models~\cite{Cirelli:2005uq}. A detailed analysis of this mass range was given  in~\cite{Hambye:2009pw}. They found  that to satisfy the WMAP constraint, a small mass splittings between the components of the inert
doublet is required, $\Delta M <20$ GeV. In this case, dark matter annihilates mainly into
$W^+W^-$, $Z^0Z^0$ and $hh$. 
%Coannihilations with $A^0$ and $H^\pm$ play an important role in the determination of the relic density.
Previous analysis, including prospects for
dark matter detection, were presented in
~\cite{Cirelli:2005uq,LopezHonorez:2006gr,Andreas:2009hj,Nezri:2009jd,Arina:2009um}.   
A more recent study showed that the high mass regime of the inert
doublet could be responsible for the DAMA signal 
through inelastic
scattering~\cite{Arina:2009um}.

The intermediate mass regime, $\mH\lesssim M_W$, is a particular feature of  the inert
doublet model which can not be reproduced in any other of the SU$(2)_L$
minimal dark matter scalar models. For these masses, dark matter annihilations are dominated by $b\bar b$ and by the three-body final state $WW^*$, and  the correct relic abundance  can be obtained with or without coannihilations between
$H^0$ and $A^0$ (collider constraints on new charged particles
prevent coannihilations with $H^\pm$). Given the richness and the simplicity of
the physics involved, various studies have been devoted to the
inert doublet model in this precise mass range. The first  analysis of this parameter space were carried out in~\cite{Barbieri:2006dq, LopezHonorez:2006gr}. Further on, it was
shown that the annihilation of inert Higgs particles could give rise to significant gamma ray
lines~\cite{Gustafsson:2007pc}. 
In \cite{LopezHonorez:2006gr,Gustafsson:2007pc,Andreas:2009hj,Nezri:2009jd,Agrawal:2008xz}, the prospects for direct and the indirect detection of dark matter were considered. Constraints
from LEPII data were derived in~\cite{Lundstrom:2008ai},  
whereas prospects for detection at colliders were analyzed in~\cite{Cao:2007rm,Dolle:2009ft,Miao:2010rg}. Recently, it was demonstrated that dark matter annihilation into the three-body final state $WW^*$ plays a crucial role in this regime, modifying in a significant way the viable parameter space of the model and the prospects for its detection~\cite{Honorez:2010re}.

Notice, in particular, that none of these known viable regions includes the
mass range $M_W<\mH<500~\gev$. It is our goal to demonstrate that the
lower part of that range, up to about $160~\gev$, is  another
viable region of the inert doublet model. Along this new viable
region, cancellations between the different diagrams that contribute
to dark matter annihilation into gauge bosons are required to satisfy
the relic density constraint. 
%
%
%
%
%L

\section{Cancellations in the $\mH>M_W$ regime}
\label{sec:canc}

\subsection{Analytical Results}

\begin{figure}[tb]
\begin{center} 
\begin{tabular}{ccc}
\includegraphics[scale=0.6]{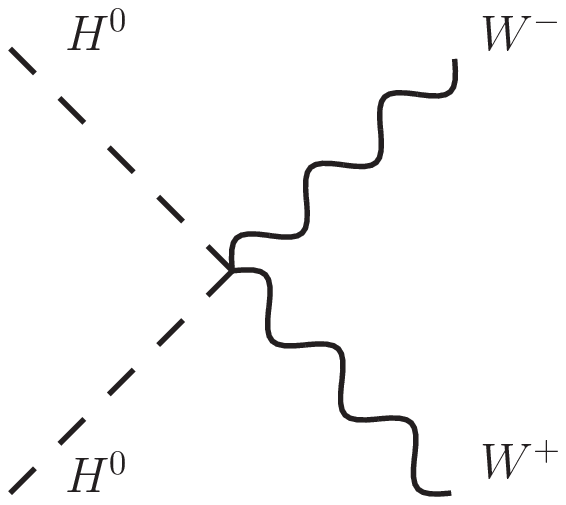} & \includegraphics[scale=0.6]{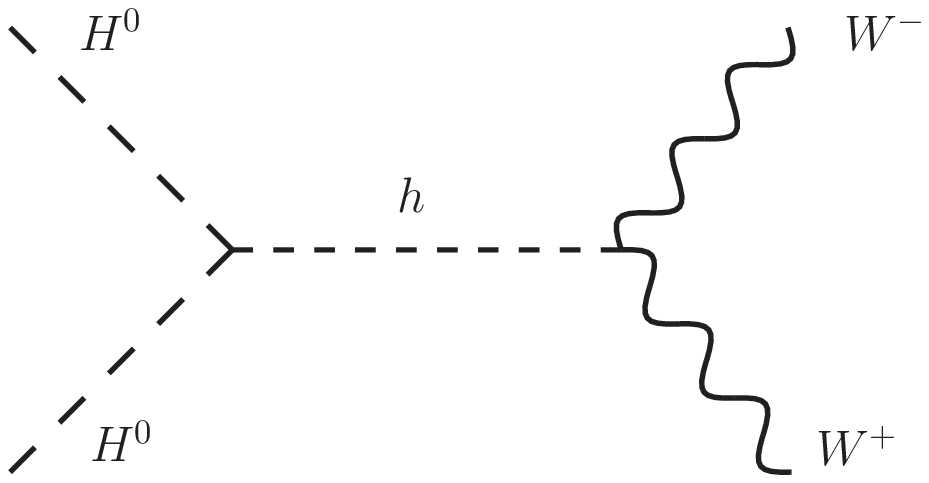} & \includegraphics[scale=0.66]{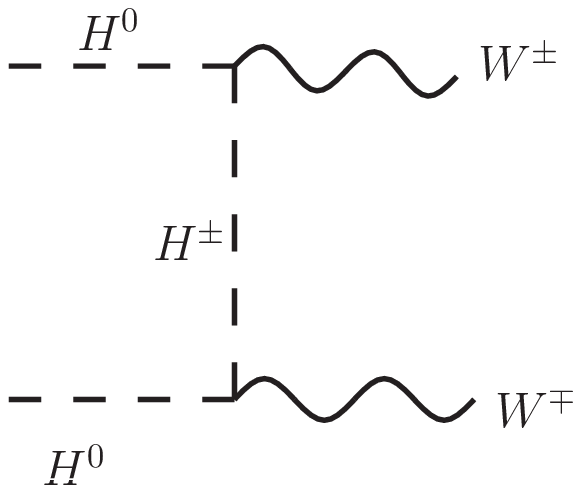}\\
(a) & (b) & (c)
\end{tabular}
\caption{Feynman diagrams contributing to dark matter annihilation into $W^+W^-$ in the inert doublet model. The exchange diagram obtained from (c) is not shown. Analogous diagrams, with $W^\pm$ replaced by $Z^0$ and $H^\pm$ by $A^0$, contribute to the annihilation into $Z^0Z^0$.\label{fig:feynman}}
\end{center}
\end{figure}
Above the $W$ threshold,  $\mH>M_W$, dark matter annihilation  into $W^+W^-$ 
becomes kinetimatically allowed, with the result that the total annihilation cross section tends to be rather large. If we estimate it from the pure gauge diagram --figure~\ref{fig:feynman}, (a)--, we get that
\begin{equation}
  \sigma v_{WW}^{\rm(s-wave)}= \frac{g^4}{128 \pi\, \mH^3 } \sqrt{
    \mH^2 - M_W^2} \left(3 + \frac{4 \mH^2 (\mH^2 - M_W^2)}{
  M_W^4}\right)\,,
\label{eq:svps}
\end{equation}
which, for $\mH\sim 100~\gev$, gives about $2\times 10^{-24}\mathrm{cm^3/s}$. And a similar number is found for the annihilation into $Z^0Z^0$. These annihilation cross sections are both much larger than those required to obtain the correct relic density ($\sigma\mathrm{v}\sim 3\times 10^{-26}\mathrm{cm^3/s}$), so it seems difficult to  satisfy the dark matter  constraint in this mass range. In fact, previous analysis have claimed that it is not possible to do so. In  \cite{Barbieri:2006dq}, they stated that the inert Higgs could only provide a subdominant component of the dark matter, while in \cite{LopezHonorez:2006gr} they found the relic density to be too suppressed to be consistent with WMAP unless $\mH>600~\gev$.

Here, we show that those conclusions are incorrect. Taking advantage of the interference
among the different diagrams that contribute to dark matter annihilation into gauge bosons, it is possible to   satisfy the relic density constraint for  $\mH\gtrsim M_W$. Thus, opening up a new viable region of the inert doublet model.

Let us now explain how these interference effects come about. The diagrams that contribute to dark matter annihilation into $W^+W^-$ (or $Z^0Z^0$) are shown in figure \ref{fig:feynman}. Their amplitudes are given by 
 \begin{eqnarray}
   i{\cal M}_p&=&i\frac{g_V^2}{2} \epsilon_{\mu}^*(p_3) \epsilon^{*\mu}(p_4)\label{eq:Mp}\\
 i{\cal M}_s&=&i\frac{\lambda_L v^2 g_V^2}{s-M_h^2-i M_h\Gamma_h} \epsilon_{\mu}^*(p_3) \epsilon^{*\mu}(p_4)\label{eq:Ms}\\
  i{\cal M}_t&=&i\frac{g_V^2}{t-M_{\chi}^2}
    p_1^\mu\epsilon_{\mu}^*(p_3)\,
    p_2^\nu\epsilon^*_\nu(p_4)\label{eq:Mt}\\
 i{\cal M}_u&=&i\frac{g_V^2}{u-M_{\chi}^2} p_2^\mu\epsilon_{\mu}^*(p_3)\, p_1^\nu\epsilon^*_\nu(p_4)\label{eq:Mu}
\end{eqnarray}
where $\epsilon_{\mu}^*(p)$ is the polarization vector of
the gauge boson with momentum  $p$, $p_1$ and $p_2$ are the momenta of the  $H^0$'s, and   $s$, $t$ and
$u$ are the Mandelstam  variables. In these equations, $g_V=g, g/\cos
\theta_{W}$ and  $\chi=H^\pm,A^0$ respectively for the
annihilations into  $V=W^\pm, Z^0$. Typically, the first two contributions (the point-like interaction and the Higgs-mediated diagram) dominate the total annihilation cross section, so we will focus on those in our analytical discussion. Far from the Higgs resonance, their squared amplitude can be written as
\begin{equation}
  |i{\cal M}_p+ i{\cal
  M}_s|^2= \frac{g^4_V}{4}\left(2-\frac14 \left(\frac{s-2M_V^2}{M_V^2}\right)^2\right)\frac{[(s-M_h^2)+2\lambda_L v^2]^2}{{(s-M_h^2)^2}} \,.
\label{eq:canc}
\end{equation}
Since  $s\simeq 4 \mH^2$ at low velocities, a cancellation between the two contributions occurs for 
\begin{equation}
\lambda_L\approx-2(\mH^2-(M_h/2)^2)/v^2\,,
\label{eq:lcanc}
\end{equation}
to which we refer as the cancellation condition. 
Hence, cancellations take place   for $\lambda_L>0$ if $\mH<M_h/2$  and for  $\lambda_L<0$  if $\mH>M_h/2$.
For example, in the case of $\mH=120$ GeV and  $\mH<M_h/2$,
 cancellations occur for $\lambda_L\simeq (0.04, 0.5,1.2,2.0)$ when $M_h \,=(250,350,450,550)$
respectively, while  for $\mH>M_h/2$,
cancellations should occur for $\lambda_L\sim [-0.35,-0.1]$ when  $M_h
\,\in \, [121,200]$. Notice also, that the cancellation condition is the same for both final states, $W^+W^-$ and $Z^0Z^0$, indicating that both processes will be simultaneously suppressed. The idea, then, is that in regions of the parameter space where the cancellation condition is satisfied, the annihilation rate could be much smaller and, accordingly, a larger relic density will be found. Perhaps, large enough to be compatible with WMAP.

The existence of cancellations in the annihilation cross section into gauge bosons had  already been 
noticed in~\cite{Lundstrom:2008ai} and  in \cite{Honorez:2010re}. But the possibility of using them to obtain viable models above $M_W$ had not been considered before. 

Let us also emphasize that the $t$- and
$u$-channels that we have neglected in the above analysis will certainly affect
the  annihilation cross section into $W^+W^-$ and $Z^0Z^0$. We will see that  in order to comply with the relic density constraint it is necessary to suppress  their contributions by considering large values of  $M_{A_0}$ and  $M_{H^\pm}$.

Suppressing  the annihilations into $W^+W^-$ and $Z^0Z^0$ is not enough to ensure a small  dark matter annihilation cross section. In addition, we must make sure that the other relevant annihilation final states, that is $hh$ and $t\bar t$, do not give a large contribution to it. If $\mH>\mh$, the total
s-wave contribution to the $H_0H_0\rightarrow hh$ annihilation process is given by
\begin{equation}
  \sigma v_{hh}^{\rm(s-wave)}= \frac{\lambda_L^2}{4\pi\,\mH^3 }\sqrt{  
  \mH^2-M_h^2} \frac{(M_h^4 - 4 \mH^4 - 2 M_h^2 v^2 \lambda_L + 
   8 \mH^2 v^2 \lambda_L)^2}{(M_h^4 - 6 M_h^2 \mH^2 + 
   8 \mH^4)^2 }\,.
  \label{eq:svhh}
\end{equation}
Using $\mH\gtrsim M_h$ and the $\lambda_L$ of equation~(\ref{eq:lcanc}),
we obtain that  $\sigma v_{hh}^{\rm(s-wave)}$ is typically of ${\cal
O} (10^{-23}) \mathrm{cm^3/s}$, giving rise to a  very suppressed relic
abundance. To be compatible with the observed dark matter density, then, the mass of the dark matter particle must be smaller than the Higgs mass ($\mH<\mh$), so that the annihilation channel into $hh$ is not kinematically allowed. This limitation of the parameter space is not that strong in the inert doublet model, for the Higgs mass can be quite large. A very stringent constraint comes instead from the annihilation into $t\bar t$.  The total s-wave contribution to the $H_0H_0\rightarrow t\bar t$ process is given by
\begin{equation}
 \sigma v_{t\bar t}^{\rm(s-wave)}=\frac{N_c \lambda_L^2 m_t^2}{\mH^3 \pi} \frac{( \mH^2-m_t^2 )^{3/2}}{ (4 \mH^2-M_h^2 )}\,,
   \label{eq:svtt}
 \end{equation}
where $m_t$ is the top mass and $N_c=3$ is the color factor. Again we find that   the $\lh$ satisfying the cancellation condition yields a very small
relic abundance for $\mH\gtrsim m_t$. Therefore, viable models must feature $\mH<m_t$. In~\cite{Hambye:2009pw}, for instance,  it was assumed that $\mH\gg m_t,\mh$, so the effect of cancellations was overlooked in their study. As a result, the minimal value of $\mH$ consistent with WMAP was found to be $535~\gev$.   From our  analysis of the mass range $M_W<\mH<500~\gev$, we conclude that models satisfying the relic density constraint could be found only along the lower part of it, specifically for  $\mH< m_t,\mh$. 

\begin{figure}[tb]
\begin{center} 
\includegraphics[scale=0.35]{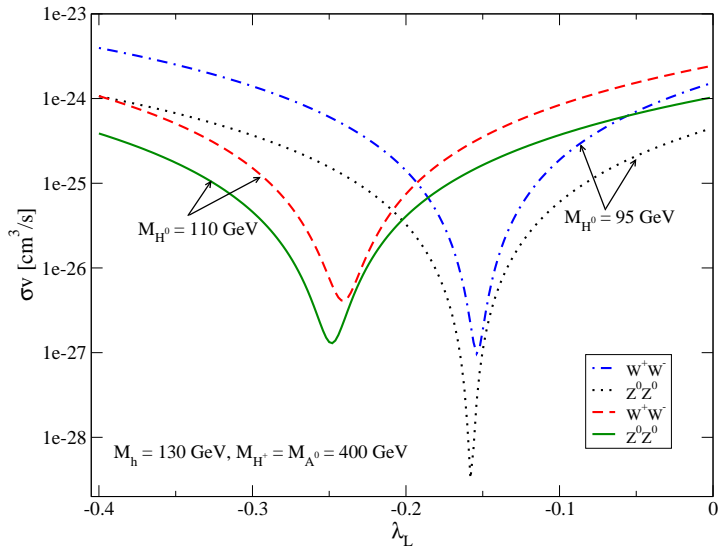}
\caption{The dark matter annihilation cross section (at low velocity)  into $W^+W^-$ and $Z^0Z^0$ as a function of $\lh$ for two different values of $\mH$: $95~\gev$ and $110~\gev$. In this figure $\mA=\mC=400~\gev$ and $\mh=130~\gev$. Notice that cancellations take place simultaneously in both channels, $W^+W^-$ and $Z^0Z^0$.\label{fig:examplesv1}}
\end{center}
\end{figure}

When evaluating   numerically the annihilation cross sections and the relic density, we always take into account all the  relevant processes, and all the diagrams that contribute to a given process. The analytical expressions derived in this section are not used for that purpose. We will verify, nevertheless, that the viable models always satisfy the cancellation condition we have obtained.

These cancellation effects we have described are illustrated in figure
\ref{fig:examplesv1}, which shows the dark matter annihilation cross
sections into $W^+W^-$ and $Z^0Z^0$ as a function of $\lh$. For this
figure we set $\mh=130~\gev$, $\mC=\mA=400~\gev$, and we consider two
different values of $\mH$: $95~\gev$ and $110~\gev$.
Equation~(\ref{eq:lcanc}) tell us that the cancellations should take
place for for negative values of $\lambda_L=-0.16$ and $-0.26$
respectively which is in good agreement with the full numerical treatment of the annihilation cross section. We  observe that the larger $\mH$ the larger the value of $|\lh|$ required to obtain cancellations.  Notice from the figure that cancellations may in fact reduce significantly the annihilation cross section and that this reduction takes place simultaneously in both channels, $W^+W^-$ and $Z^0Z^0$.

Since $\oh\propto 1/\sigma \mathrm{v}$,  we expect these cancellations in the annihilation cross section to  increase significantly the inert Higgs relic density, opening up the possibility of finding viable models for $\mH\gtrsim M_W$. Next, we study, with several examples,  the dependence of the relic density on the parameters of the models and show that there are indeed viable models in this region of the parameter space.

\subsection{Examples}
\begin{figure}[tb]
\begin{center} 
\includegraphics[scale=0.35]{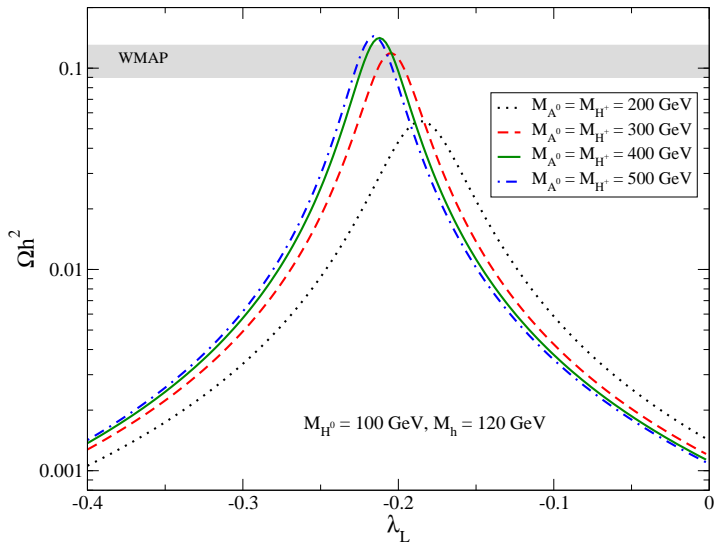}
\caption{The relic density as a function of $\lh$ for several values of $\mA=\mC$: $200$, $300$, $400$, $500~\gev$.  In this figure $\mH=100~\gev$ and $\mh=120~\gev$.\label{fig:example}}
\end{center}
\end{figure}
Figure \ref{fig:example} shows the relic density as a function of $\lh$ for $\mH=100$ GeV, $\mh=120$ GeV and different values of $\mA$ and $\mC$. For simplicity we consider models with $\mA=\mC$ and we let them vary between $200~\gev$ and $500~\gev$. The horizontal band shows the region compatible with the observed dark matter density, $0.09<\oh<0.13$. The effect of cancellations on  the relic density is apparent in this figure. They  increase $\oh$ significantly as $\lh$ approaches about $-0.2$, where the relic density reaches its  maximum value. This value depends slightly on the value of $\mA=\mC$ but is in any case close to  $0.1$. Far from this maximum, say for $\lh=-0.4$ or $\lh=0$, the relic density is two orders of magnitude smaller. From this figure we also learn that cancellations are necessary but not sufficient to ensure a relic density compatible with the observations. In addition, the mass of the charged scalar (and of the CP-odd scalar) should be large enough to suppress the $H^\pm$ (and $A^0$) mediated contributions. If, for instance, $\mC\lesssim 200~\gev$  cancellations are still present but the  relic density is always below the observed range. For the other values of $\mC$ we consider, $300, 400, 500~\gev$, it is always possible to satisfy the dark matter constraint for a certain range of $\lh$. And the allowed values of $\lh$ depend only slightly on the precise value of $\mC=\mA$. This first example demonstrates that in the inert doublet model it is indeed possible to satisfy the relic density constraint for $\mH\gtrsim M_W$.

\begin{figure}[tb]
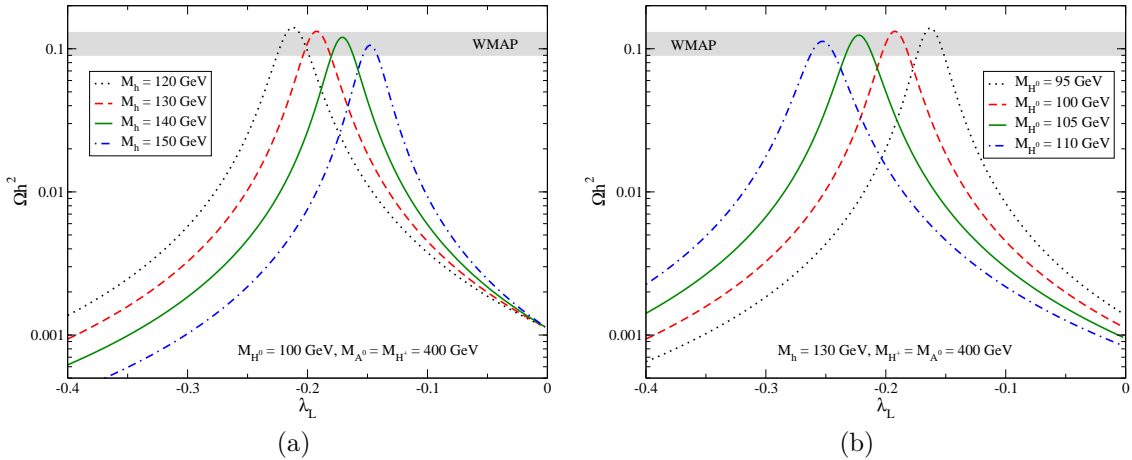

\begin{center} 
\begin{tabular}{cc}
\hspace{-4mm}\includegraphics[scale=0.27]{exampleb.eps} & \includegraphics[scale=0.27]{examplef.eps}\\
(a) & (b)
\end{tabular}
\caption{(a) The relic density as a function of $\lh$ for different values of $\mh$. For this figure we set $\mH=100~\gev$ and $\mA=\mC=400~\gev$. (b) The relic density as a function of $\lh$ for different values of $\mH$. The other parameters were taken as $\mh=130~\gev$ and $\mA=\mC=400~\gev$.\label{fig:exampleb}}
\end{center}
\end{figure}

The dependence of the relic density on $\mH$ and $\mh$ is shown in figure \ref{fig:exampleb}. They  both display the relic density as a function of $\lh$ for  $\mA=\mC=400~\gev$.  In (a), $\mH$ was set to $100~\gev$ and  four different values of $\mh$ ($120,130,140,150~\gev$) were considered.  As before, the horizontal band shows the region compatible with the observed dark matter density. The effect of cancellations are evident: they allow to satisfy the dark matter bound for $\lh$ approximately between $-0.15$ (for $\mh=150~\gev$) and $-0.22$ (for $\mh=120~\gev$).  As expected from equation (\ref{eq:lcanc}), the value of $\lh$ that gives the maximum relic density moves toward less negative values as $\mh$ is increased. Because  the Higgs-mediated  contribution to dark matter annihilation is proportional to $\lh$,  the Higgs-dependence of the relic density disappears for $\lh=0$, as observed in the figure. At that point, the relic density is driven by the direct annihilation into gauge bosons (figure \ref{fig:feynman},(a)) resulting in $\oh\sim 10^{-3}$.  We can conclude from this figure that for $\mH=100~\gev$ and $\mA=\mC=400~\gev$ there is a significant range of Higgs masses for which the relic density can be consistent with the observations.

In figure \ref{fig:exampleb} (b) $\mh$ was set to $130~\gev$ and four different values of $\mH$ ($95,100,105,110~\gev$) were considered. We see that the position of the peak moves toward more negative values of $\lh$ as $\mH$ is increased, in agreement with the results found previously. To obtain the observed relic density, values of $\lh$ approximately between $-0.15$ (for $\mH=95~\gev$) and $-0.27$ (for $\mH=110~\gev$) are required. These figures illustrate that the new viable region occupies a non-negligible volume in the parameter space of the inert doublet model, spanning a significant range in $\mH$, $\lh$, $\mh$, $\mA$ and $\mC$.

\begin{figure}[tb]
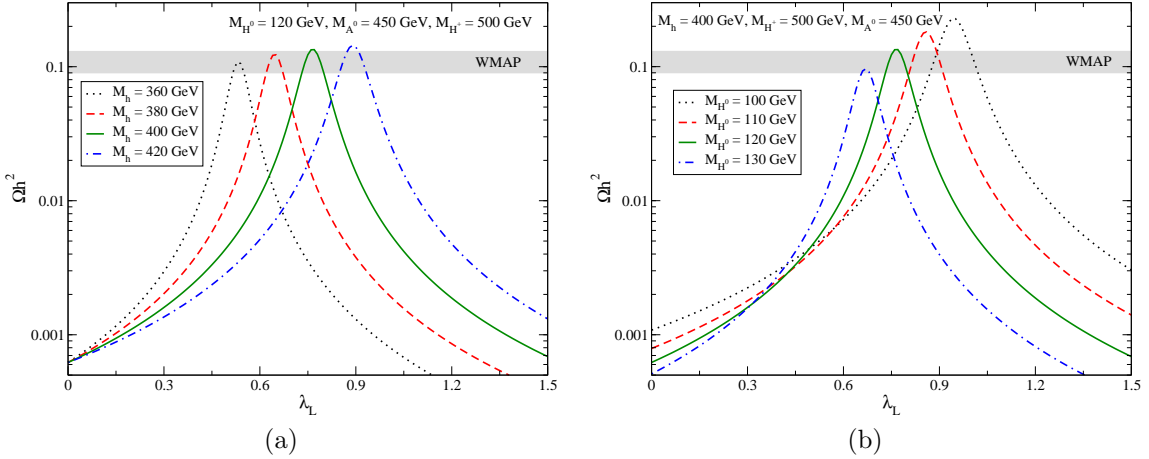

\begin{center} 
\begin{tabular}{cc}
\includegraphics[scale=0.27]{examplee.eps} & \includegraphics[scale=0.27]{exampled.eps}\\
(a) & (b)
\end{tabular}
\caption{(a) The relic density as a function of $\lh$ for several Higgs masses. In this figure we set $\mH=120~\gev$, $\mA=450~\gev$ and $\mC=500~\gev$. (b) The relic density as a function of $\lh$ for  different values of $\mH$: $100$, $110$, $120$, $130~\gev$. The other parameters were taken as $\mh=400~\gev$, $\mA=450~\gev$ and $\mC=500~\gev$.  \label{fig:exampled}}
\end{center}
\end{figure}

So far we have illustrated the effect of cancellations on the relic density only for the light Higgs regime, $\mh\lesssim 200~\gev$. A remarkable feature of the inert doublet model is that it allows for a much heavier Higgs without contradicting electroweak precision data. As we will see, cancellations leading to  viable models with $\mH\gtrsim M_W$ can also take place in the  heavy Higgs regime of the model, that is for $\mh>200~\gev$.

Figure \ref{fig:exampled} is analogous to figure \ref{fig:exampleb}
but for a heavy Higgs boson, $\mh\sim 400~\gev$. In both panels $\mC$
was set to $500~\gev$ and $\mA$ to $450~\gev$ (they must now give a
positive contribution to $\Delta T$). In (a)  the relic density is
shown as a function of $\lh$ for  $\mH=120~\gev$ and four different
values of $\mh$: $360, 380,400,420~\gev$. First of all, notice that in
this case cancellations occur for positive rather than negative values
of $\lh$. As expected from the cancellation condition, the position of the peak  moves toward larger values of $\lh$ as $\mh$ is increased. The values of $\lh$ that give the correct relic density are between $0.5$ (for $\mh=360~\gev$) and about $0.9$ (for $\mh=420~\gev$). For $\lh=0$, $\oh$  becomes independent of $\mh$, with a value of about $6\times 10^{-4}$.  In (b)  the relic density is shown as a function of $\lh$ for  $\mh=400~\gev$ and four different values of $\mH$: $100, 110,120,130~\gev$. The viable region corresponds in this case to $\lh$ between $0.7$ (for $\mH=130~\gev$) and $1.0$ (for $\mH=100~\gev$). We observe that the maximum value of $\oh$ increases as  $\mH$ is decreased.  In fact, if $\mH=130~\gev$ the maximum relic density is as large as the WMAP lower limit. So, it is not possible to  satisfy the relic density constraint (keeping the other parameters fixed) for $\mH>130~\gev$. But if $\mH=100~\gev$ the relic density can be larger than the WMAP upper limit, reaching about $0.23$. As a result, it is easier to satisfy the relic density constraint with a lighter inert Higgs.   

As we have seen in the previous figures, cancellations between the different diagrams that contribute to dark matter annihilation into gauge bosons indeed allow to satisfy the relic density constraint for inert Higgs masses above $M_W$ (and below $m_t$). In other words, there certainly exist regions in the viable parameter space of the inert doublet model that had not been considered before. What we are lacking at this point is a more complete characterization of this new viable region. We would like to know, for instance, what is the largest dark matter mass ($\mH$) that is allowed, what are the typical values of $\lh$, $\mh$, $\mC$ and $\mA$, or what is the most promising strategy to detect the inert Higgs along this region. In the next sections we will address precisely these issues.     
\section{The new viable region}
\label{sec:new}
So far, we have demonstrated the existence of a new viable region in the inert doublet model. This previously overlooked region features dark matter masses larger than $M_W$ and is consistent with the relic density constraint thanks to cancellations between the Higgs mediated diagram and the direct annihilation diagrams into $W^+W^-$ and $Z^0Z^0$. In this section we perform a systematic analysis of this new viable region of the parameter space. To that end, we make use of  Markov Chain Monte Carlo (MCMC) techniques to efficiently explore the full parameter space of the inert doublet model and find a large number of viable models in the region of interest to us. Then, we proceed to analyze such models in detail. 
\subsection{The scan}
The inert doublet model contains $5$ free parameters that can be taken to be $\mH$, $\mC$, $\mA$, $\lh$,  and $\mh$. We allow these parameters to vary within the following ranges:
\begin{eqnarray}
80~\gev<&\mH&<200~\gev\\
80~\gev<&\mC&<1000~\gev\\
80~\gev<&\mA&<1000~\gev\\
-12<&\lh&<12\\
114~\gev<&\mh&<600~\gev
\end{eqnarray}
MCMC tecniques allow us to scan freely and efficiently over these $5$
parameters. Let us emphasize that these techniques have already been
successfully used to scan and analyze the viable parameter space of
other models of dark matter, see \cite{Baltz:2004aw,deAustri:2006pe}. Here we closely follow the procedure outlined in \cite{Baltz:2004aw}. 

After scanning the above parameter space, we get a large sample of
viable models,  satisfying all the experimental and theoretical bounds and the relic density constraint. For our analysis we generated about 40000 such models within the new viable region. In the remaining part of the paper, we will work exclusively with this sample of models, analyzing its main features and studying its implications. 

\subsection{Different projections of the viable region}
\label{sec:diff-proj-viable}

\begin{figure}[t]
\begin{center} 
\includegraphics[scale=0.35]{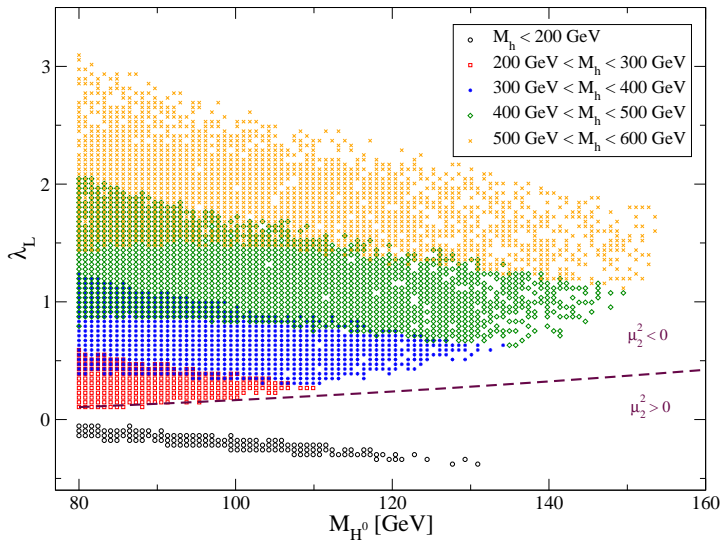}
\caption{The new viable region projected on the plane ($\mH$,$\lh$). The different symbols (and colors) distinguish among the possible ranges of $\mh$. The dashed line is the contour $\mutwo=0$. Above that line $\mutwo$ is negative whereas it is positive below it. \label{fig:mh0lamh0}}
\end{center}
\end{figure}

To facilitate the analysis and visualization of the 5-dimensional
region where the viable models are located, we will  project it
onto different two-dimensional planes. Figure \ref{fig:mh0lamh0}, for
instance, shows the new viable region of the inert doublet model in
the plane ($\mH$,$\lh$). To generate this kind of figures, we created
a suitable grid in the two variables ($\mH$ and $\lh$ in this case)
and we place a point at a given value of them if there is at least one
viable model in our sample with such coordinates. Sometimes we also
differentiate (using symbols and colours) various ranges for other
important parameters (the Higgs mass in figure \ref{fig:mh0lamh0})
that may provide additional information.  Notice in figure
\ref{fig:mh0lamh0} that the value of $\mH$ over the new region extends
from $M_W$ up to about $160~\gev$. We argued in section \ref{sec:canc}
that $\mH$ had to be smaller than $m_t=171.4~\gev$ but at that point
we did not know how close to $m_t$ it could get. In $\lh$, the range of
possible values  extends from $-0.5$ to about $3.0$.  A sharp contrast
is observed between the models with $\lh<0$ and those with $\lh>0$. In
the former case ($\lh<0$), the maximum value of $\mH$ is about
$130~\gev$ and the Higgs boson is necessarily light, $\mh<200~\gev$, as
cancellations occur for $\mH>M_h/2$. In the latter case ($\lh>0$), the maximum value of $\mH$ can be larger, strongly depending on the range of Higgs masses. Thus, for $200~\gev<\mh<300~\gev$ the maximum $\mH$ is only about $110~\gev$ whereas it can reach almost $160~\gev$ for  the largest Higgs masses, $\mh>500~\gev$. It is also apparent from the figure, that for $\lh>0$ the larger the Higgs mass the larger $\lh$ must be. Given that cancellations are needed to obtain the correct relic density, it is not surprising that no viable models are found with $\lh$ close to zero, as in that region the Higgs-mediated diagram becomes negligible. Finally,  the dashed line shows the contour $\mutwo=0$. From the figure we see that viable points with $\mutwo>0$ feature $\lh<0$ and viceversa.

\begin{figure}[tb]
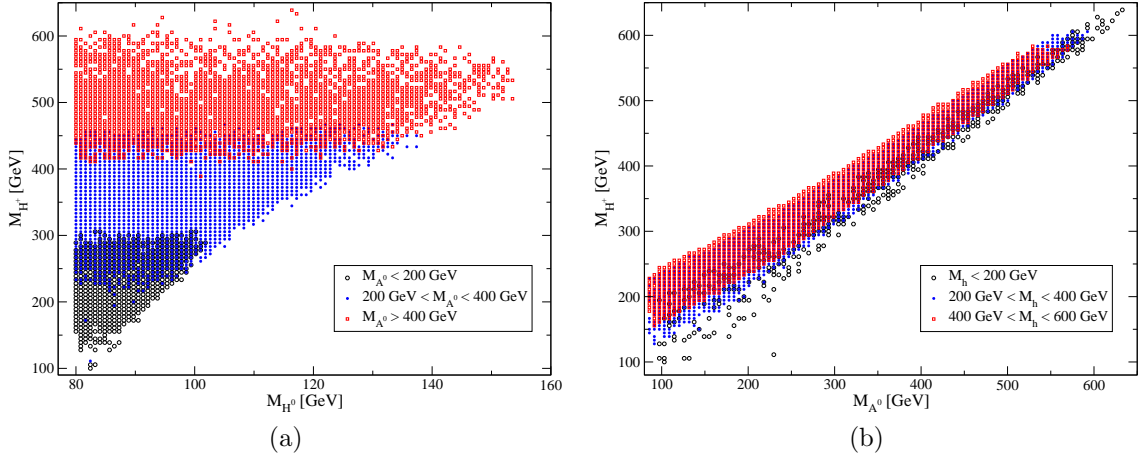

\begin{center} 
\begin{tabular}{cc}
\includegraphics[scale=0.27]{mh0mhc.eps} & \includegraphics[scale=0.27]{ma0mhc.eps}\\
(a) & (b)
\end{tabular}
\caption{(a) The new viable region projected on the plane ($\mH$,$\mC$). Different symbols (and colors) are used to distinguish the possible values of $\mA$. (b) The new viable region projected on the plane ($\mA$,$\mC$). Three possible ranges of $\mh$ are set apart with different symbols (and colors).\label{fig:mh0mhc}}
\end{center}
\end{figure}

The dependence of the new viable region on $\mC$ and $\mA$ is
illustrated in figure \ref{fig:mh0mhc}. Panel (a) shows the projection
on  the plane ($\mH$, $\mC$) for different ranges of $\mA$. The trend
is clear: the heavier the dark matter particle the larger $\mC$ must
be. If $\mH<100~\gev$, $\mC$ can take values within a wide range, but
as $\mH$ increases, higher and higher values of $\mC$ are
required. If, for instance, $\mH>130~\gev$, $\mC$ should be larger
than about $400~\gev$. And the viable models with the highest dark
matter masses ($\mH>150~\gev$) all feature a heavy $H^\pm$ (and a
heavy $A^0$), $\mC>500~\gev$. It is also observed in the figure that
there is a correlation between the values of $\mC$ and $\mA$. Large
values of $\mC$ are associated with large values of $\mA$.  This
correlation is illustrated in more detail in panel (b) of the same
figure, which shows the viable region in the plane ($\mA$,$\mC$) for
different ranges of the Higgs mass. In this plane, the viable models
are concentrated along a band just above the $\mC=\mA$ line (the
diagonal). In other words, most models feature $\mC>\mA$ but their
mass splitting is never that large. Let us emphasize that this
condition is not enforced by the dark matter constraint but rather by
electroweak precision data, see equations~(\ref{deltat}) and (\ref{eq:ewpd}).  We see then that both $\mA$ and $\mC$ can be as large as $650~\gev$ and that the models with the highest dark matter masses ($\mH\gtrsim 140~\gev$)  require them to be larger than about $500~\gev$.

\begin{figure}[t]
\begin{center} 
\includegraphics[scale=0.35]{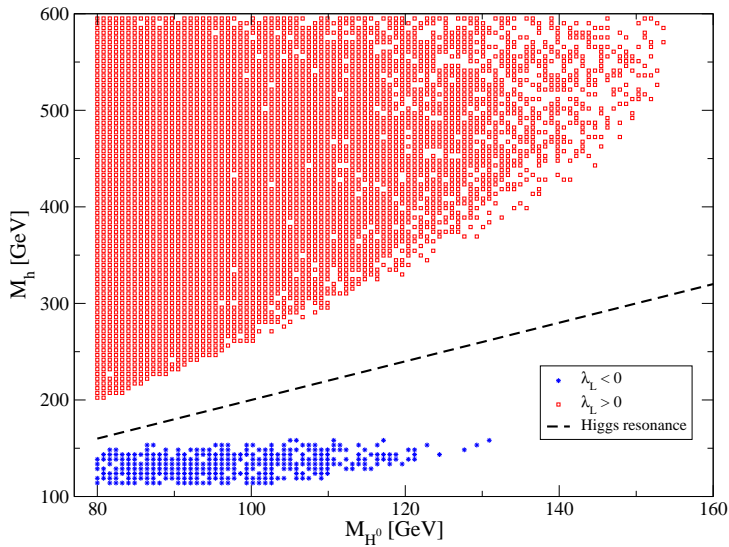}
\caption{The new viable region projected on the plane
  ($\mH$,$\mh$). The line shows the position of the Higgs resonance
  region, $2\mH=\mh$.  Different colors and symbols are used to differentiate models with $\lh>0$ from those with $\lh<0$.\label{fig:mh0mh}}
\end{center}
\end{figure}

In figure \ref{fig:mh0mh} we show the new viable region in the plane ($\mH$,$\mh$). The Higgs resonance line, $2\mH=\mh$, is also shown for illustration. A priori we do not expect to find viable models close to that line, for resonant annihilations would drive the relic density to very small values.  Points lying below the Higgs resonance line correspond to models with $2\mH>\mh$ and they all feature $\lh<0$. From the figure we see that  such points extend up to about $130~\gev$ in $\mH$ and about $160~\gev$ in $\mh$. They correspond, therefore, to what we call the light Higgs regime. Points above the Higgs resonance line ($2\mH<\mh$), in constrast, all feature a positive value of $\lh$. It can be observed that they include dark matter masses as high as $160~\gev$ and Higgs bosons heavier than $200~\gev$ and extending up to $600~\gev$. Hence, they lie within the heavy Higgs regime. Finally it is worth mentioning that no viable models are found  for  Higgs masses approximately between $160~\gev$ and  $200~\gev$.

\begin{figure}[t]
\begin{center} 
\includegraphics[scale=0.35]{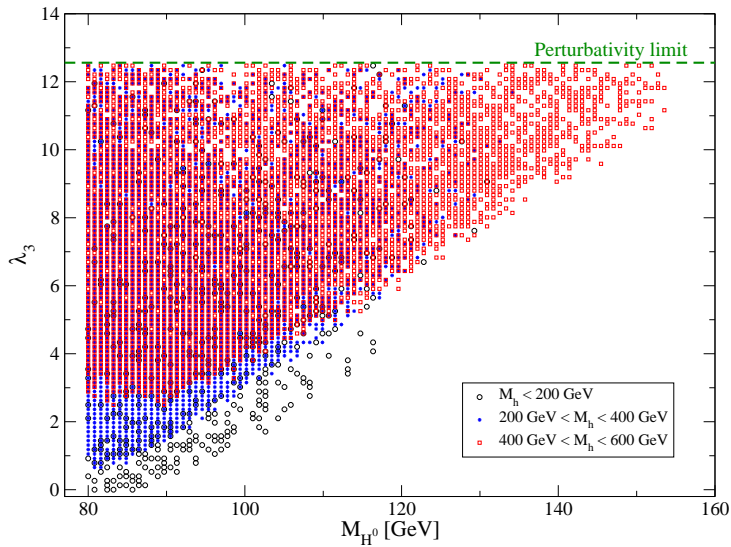}
\caption{The new viable region projected on the plane ($\mH$,$\lambda_3$). The different symbols (and colors) distinguish among the possible ranges of $\mh$. The dotted line shows the perturbativity limit, $\lambda_3=4\pi$. \label{fig:mh0lam3}}
\end{center}
\end{figure}

We have in this way determined the position of the new viable region in the 5-dimensional space formed by $\mH$, $\mA$, $\mC$, $\lh$ and $\mh$. It is also useful to understand its position in the space of $\lambda$ parameters, which consists of $\lambda_{3}$, $\lambda_{4}$ and $\lambda_5$. The reason is that some constraints, such as the perturbativity and the vacuum stability bounds, are applied directly to these couplings, not to scalar masses or to $\lh$. Since many models feature a rather heavy $H^\pm$ and $\mutwo$ negative (equivalently $\lh$ positive), we expect that a larger value of $\lambda_3$ will be required --see equation (\ref{eq:masses}). That is what we observed in figure \ref{fig:mh0lam3}. It displays the new viable region in the plane ($\mH$,$\lambda_3$) for different ranges of $\mh$. We see that $\lambda_3$ can indeed be large, with some models lying right at the perturbativity limit, $\lambda_3=4\pi$ (dotted line). In the figure we observe that the models with the largest dark matter masses ($\mH\gtrsim 140~\gev$), in particular, feature $\lambda_3\gtrsim 9$. It must be emphasized, however, that $\lambda_3$ is not that large over the entire new viable region. Models with $\lambda_3<6$, for instance, extend up  to $\mH=120~\gev$ while those with $\lambda_3\lesssim 2$ can reach up to $105~\gev$. Thus, if  one wants to impose a more restrictive perturbativity bound on $\lambda_3$, this figure can be used to find out the resulting viable range of $\mH$.

\begin{figure}[t]
\begin{center} 
\includegraphics[scale=0.55]{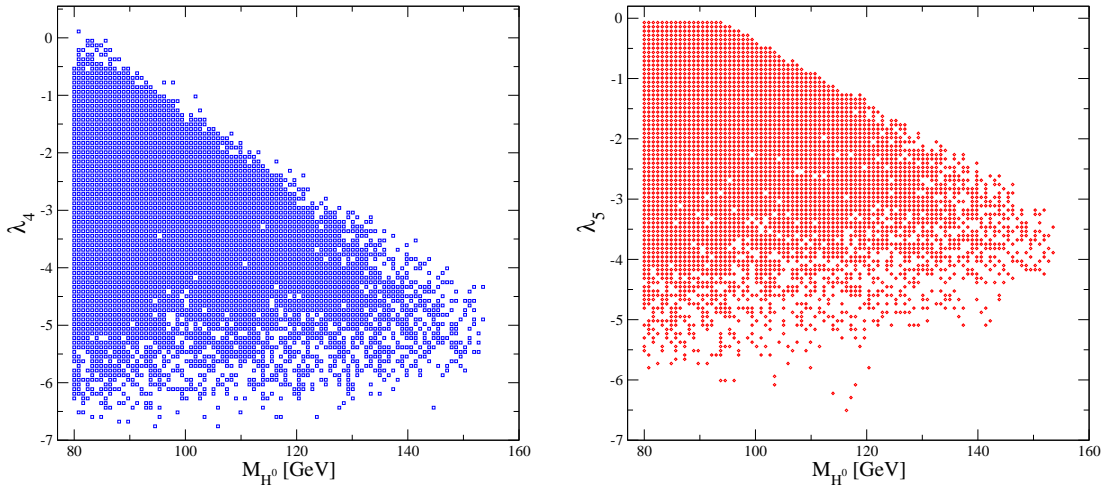}
\caption{The new viable region projected on the planes ($\mH$,$\lambda_{4}$) (left panel) and ($\mH$,$\lambda_{5}$) (right panel).\label{fig:mh0lam45}}
\end{center}
\end{figure}

Figure \ref{fig:mh0lam45} shows the new viable regions in the planes ($\mH$,$\lambda_{4}$) and ($\mH$,$\lambda_{5}$). $\lambda_4$ turns out to be negative in most models and reaches at most $|\lambda_4|=7$, significantly below the perturbativity limit (not shown). The possible values of $\lambda_4$ move toward lower values as $\mH$ is increased. Hence, for $\mH=100~\gev$ $\lambda_4\in\left(-1,-7\right)$ whereas for  $\mH=130~\gev$ $\lambda_4\in\left(-3,-7\right)$. Notice also that $\lambda_4\sim -5$ for the models with the largest values of $\mH$. $\lambda_5$, which determines the mass splitting between $H^0$ and $A^0$,  is by definition negative ($\mA>\mH$). We see, in panel (b), that its absolute value is never larger than $7$ and only rarely exceeds $5$, placing it comfortably within the perturbative regime. As with $\lambda_4$, the allowed values of $\lambda_5$ tend to move toward lower values (higher $|\lambda_5|$) as $\mH$ is increased.  For the models with the highest dark matter masses, $\lambda_5$ is about $-4.0$. Consequently, a more restrictive pertubativity bound on $\lambda_4$ and $\lambda_5$, say $\lambda_{4,5}<5$, could be imposed without affecting the viable range of $\mH$.

We thus conclude this subsection regarding the projections of the viable region onto different two-dimensional planes. Before addressing the implications of the new viable region for the direct and the indirect detection of inert Higgs dark matter, we will discuss the issue of the  fine-tuning required to obtain the correct relic density within this new viable region. Is it large, moderate or small? How does it compare against the fine-tuning found in other models of dark matter?  
\subsection{Fine-tuning}
Given that in it cancellations are necessary to obtain the right
relic density, one may think that this new viable region of the inert doublet model is disfavoured by fine-tuning arguments. Here, we quantify the fine-tuning of the viable models and show that that is not really the case. Many of them feature low to moderate fine-tuning and none of them has a very large one.

The issue of the fine-tuning required to satisfy the dark matter constraint  is not new. In \cite{Ellis:2001zk}, where this fine-tuning was first quantified,  it was proposed to use as a fine-tuning parameter  the logarithmic sensitivity of the relic density to variations in the input parameters of the model. If $a_i$ ($i=1,...,n$) are free parameters of a given dark matter model,  the fine-tuning parameter (of the relic density) with respect to $a_i$, $\Delta_{\oh,a_i} $ is given by 
\begin{equation}
\Delta_{\oh,a_i}\equiv \frac{\partial \log\oh}{\partial \log a_i}\,.
\end{equation}
And the total fine-tuning, $\Delta_{\oh,\mathrm{total}}$, is obtained by summing in quadrature the contributions of the different parameters of the model. Thus, in the inert doublet model we have that
\begin{equation}
\Delta_{\oh,\mathrm{total}}=\sqrt{\Delta_{\oh,\mH}^2+\Delta_{\oh,\mC}^2+\Delta_{\oh,\mA}+\Delta_{\oh,\lh}^2+\Delta_{\oh,\mh}}\,.
\end{equation} 
Hence, the fine-tuning parameter is large if a small variation in the parameters of the models leads to a large modification of the relic density.  If, for instance, $\ftp\lesssim 10$, a measurement of the parameters of the model at the $10\%$ level will enable to compute $\oh$  to within a factor $\mathcal{O}(2)$. As a rule of thumb, one can say that the fine-tuning is small if $\ftp\lesssim 10$, moderate for $10<\ftp\lesssim 100$ and large if $\ftp>100$. 
\begin{figure}[t]
\begin{center} 
\includegraphics[scale=0.35]{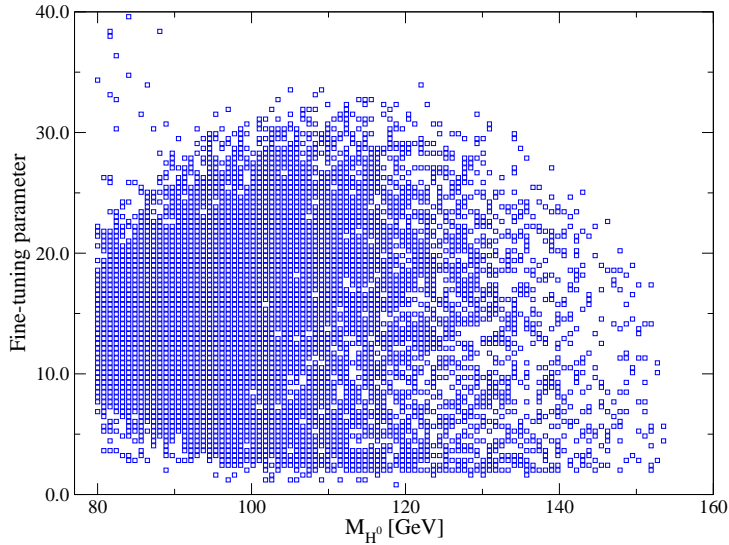}
\caption{The fine-tuning parameter, $\Delta_{\oh,\mathrm{total}}$, as a function of $\mH$ along the new viable region of the model.\label{fig:mh0ft}}
\end{center}
\end{figure}

Figure \ref{fig:mh0ft} shows the fine-tuning parameter as  a function of $\mH$ along the new viable region of the inert doublet model. $\ftp$  is typically smaller than $30$, with plenty of models featuring $\ftp<10$ over the entire range of $\mH$ and no models with a fine-tuning parameter larger than $40$. In particular, notice that there is no correlation between $\ftp$ and $\mH$. Low fine-tuning models, $\ftp\lesssim 10$, exist for $\mH$ just slightly above $M_W$ as well as for $\mH> 140~\gev$. And the same holds for moderate fine-tuning models, $\ftp\gtrsim 20$.

These results can be compared with those obtained in previous works for one of the most studied models of dark matter, the Constrained Minimal Supersymmetric Standard Model (CMSSM or mSUGRA). It turns out that  the viable regions of the CMSSM tend to be  highly fine-tuned, with some regions, such as the focus-point and the funnel region, featuring a $\ftp$ greater than $100$ \cite{Ellis:2001zk,Baer:2009vr}. In contrast, in the new viable region of the inert doublet model we are studying, the fine-tuning parameter is never that high. 

\section{Implications for detection}
\label{sec:det}

In the previous section we described in detail the new viable region of the inert doublet model. Here, we examine its implications for the detectability of the inert doublet model via collider signals and the direct and the indirect detection of inert Higgs dark matter.

\subsection{Collider signals}

\begin{figure}[tb]
\begin{center} 
\includegraphics[scale=0.35]{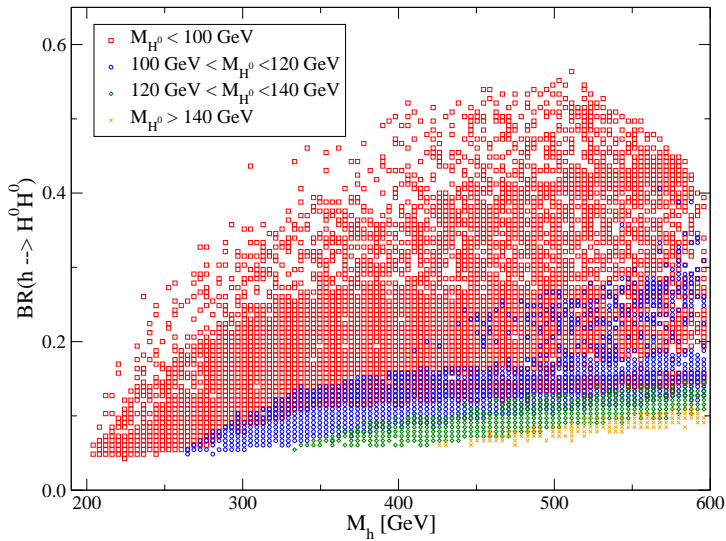}
\caption{The decay branching ratio of the Higgs boson into the inert scalars as a function of $\mh$ for different ranges of $\mH$. Only models with $\lh>0$ give a non-zero contribution to this branching.\label{fig:mh0brinv}}
\end{center}
\end{figure}

The inert doublet model can be tested at colliders via the production and decay of inert particles or through the increase in the decay width of the Higgs boson \cite{Barbieri:2006dq}. Indeed, the decay processes
\begin{equation}
  h\rightarrow H^0 H^0, A^0 A^0,H^+H^-
\end{equation}
may enhance the Standard Model Higgs decay width when
$M_h>2 \mH$, with important implications for Higgs searches at Tevatron and at the LHC. For the dark matter mass range we are studying, this increase is mainly due to the $H^0H^0$ final state and it takes place
only when $\lambda_L>0$ and $M_h>200$ GeV, see
figure~\ref{fig:mh0lamh0}. As a consequence, the Higgs decay width
considered in the recent summary of  Higgs searches at Tevatron
below 200 GeV~\cite{:2010ar} is left untouched  by the inert scalars
for $\mH>M_W$. 
The LHC, on the other hand, is expected to
perform direct measurements of the Higgs decay width with a precision
of about 6\%~\cite{Bernardi:2008zz} for $300~\gev < M_h < 700~\gev$.

Figure~\ref{fig:mh0brinv} shows the Higgs branching ratio into $H^0H^0$ as a function of the Higgs mass for our sample of viable models. It lies approximately between $5\%$ and $60\%$. In this figure, different ranges of $\mH$ are distinguished.  For models with $\mH>140~\gev$, BR($h\to H^0H^0$) is observed to be about $10\%$ or less.  Branching ratios greater than $20\%$ or so can also be found but only  for models with the smallest dark matter masses, $\mH\lesssim100~\gev$. In such case, the inert Higgs could be  within the reach of LHC via measurements of the Higgs decay width.

\subsection{Indirect detection}

\begin{figure}[tb]
\begin{center} 
\includegraphics[scale=0.35]{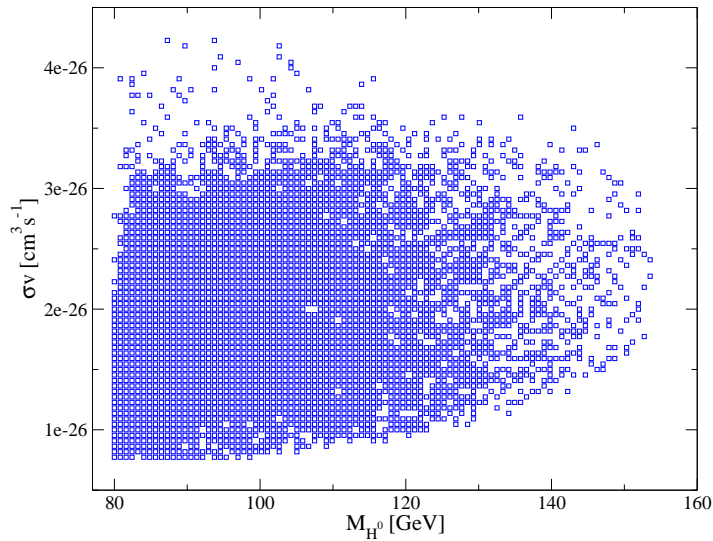}
\caption{The present dark matter annihilation rate, $\sigma\mathrm{v}$, as  a function of $\mH$ along the new viable region of the model.Notice that $\sigma\mathrm{v}$ varies only over a small range, approximately between $1\times 10^{-26}$ and $4\times 10^{-26}\mathrm{cm^3/s}$.\label{fig:mh0sv}}
\end{center}
\end{figure}

The indirect detection signatures of dark matter are determined, on
the particle physics side, by the present annihilation rate and by the
annihilation branching ratios. Figure \ref{fig:mh0sv} shows the
present dark matter annihilation rate, $\sigma\mathrm{v}$, as  a
function of $\mH$ along the new viable region ($\mathrm{v}$ is taken
to be $10^{-3}$). We see that most points
are confined within  a small interval between
$0.5\times10^{-26}\mathrm{cm^3/s}$ and $4\times
10^{-26}\mathrm{cm^3/s}$.  In other words, the present annihilation
rate is always close to the typical value for a \emph{thermal} relic,
$\langle \sigma\mathrm{v}\rangle\sim 3\times 10^{-26}\mathrm{cm^3/s}$. This outcome is a consequence of the fact that the  annihilation  of inert Higgs dark matter is dominated by its s-wave component, so $\sigma\mathrm{v}$ remains unchanged between the early Universe and the present time. Besides, coannihilations, with $A^0$ or $H^\pm$, play no important role in the region we are considering, as they would drive the relic density to even smaller values. 

\begin{figure}[tb]
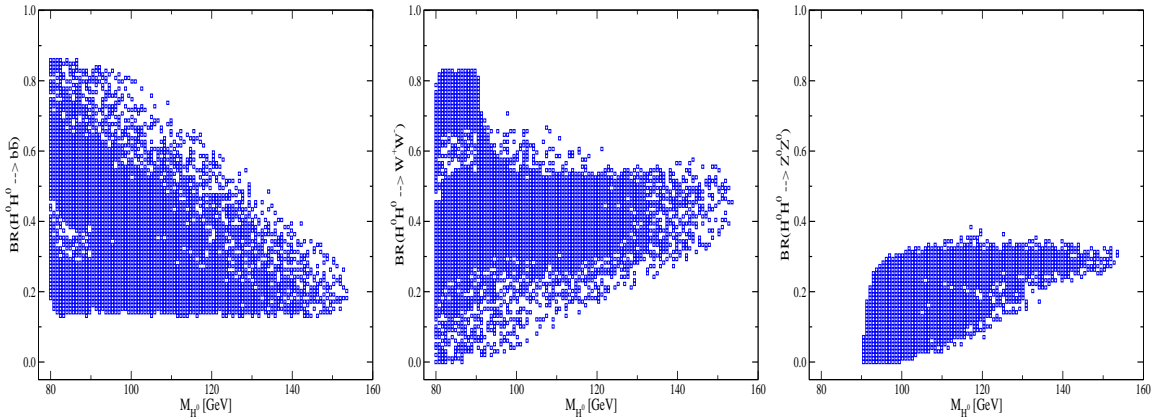

\begin{center}
\begin{tabular}{ccc} 
\hspace{-5mm}\includegraphics[width=5.0cm,height=5.5cm]{mh0brbb.eps} & \hspace{-3mm}\includegraphics[width=5.0cm,height=5.5cm]{mh0brww.eps} & \hspace{-3mm}\includegraphics[width=5.0cm,height=5.5cm]{mh0brzz.eps}
\end{tabular}
\caption{The main annihilation branching ratios of inert Higgs dark matter as a function of $\mH$. The different panels show, from left to right, $b\bar b$, $W^+W^-$, and $Z^0Z^0$. \label{fig:mh0brs}}
\end{center}
\end{figure}

In the new viable region, the inert Higgs annihilates mainly into
$b\bar b$, $W^+W^-$ and $Z^0Z^0$ (the  $hh$ and $t\bar t$ final states are
kinematically forbidden). Figure \ref{fig:mh0brs} shows the
annihilation branching ratio (BR) into these three final states as a function of $\mH$. We see, in the left panel, that the maximum value of BR($H^0H^0\to b\bar b$) decreases with $\mH$, going from about $0.85$ for $\mH\sim 80~\gev$ to about $0.2$ for $\mH\sim 140~\gev$. The minimum value of BR($H^0H^0\to b\bar b$), on the other hand, tends to remain constant at about $0.15$. In particular, the $b\bar b$ final state is never negligible within the new viable region. The center panel shows the branching ratio into $W^+W^-$. As $\mH$ crosses the $Z^0$ threshold, the annihilation into $Z^0Z^0$ start playing a significant role and the branching into $W^+W^-$ is therefore reduced. It is also observed in  figure~\ref{fig:mh0brs} that the range of variation of BR($H^0H^0\to W^+W^-$) shrinks as $\mH$ increases. For $\mH>120~\gev$ BR($H^0H^0\to W^+W^-$) lies approximately between $0.3$ and $0.5$. Finally, the right panel shows BR($H^0H^0\to Z^0Z^0$). It never exceeds $40\%$, reaching values between $0.25$ and $0.35$ for $\mH>120~\gev$, and  lower  values for smaller $\mH$. Comparing the three panels we can deduce that $b\bar b$, $W^+W^-$ and $Z^0Z^0$ are indeed the dominant annihilation channels, as their combined branching is always close to $100\%$. The remaining annihilation channels, into light fermions, typically account for less than $10\%$ of the annihilations.  

The present annihilation rate and the branching ratios that we have calculated determine the indirect detection signatures of inert Higgs dark matter. To illustrate the prospects for the indirect detection of dark matter along this new viable region, we  compute the expected gamma ray flux from the annihilation of inert Higgs particles and compare it with present data from FERMI.  

\begin{figure}[t]
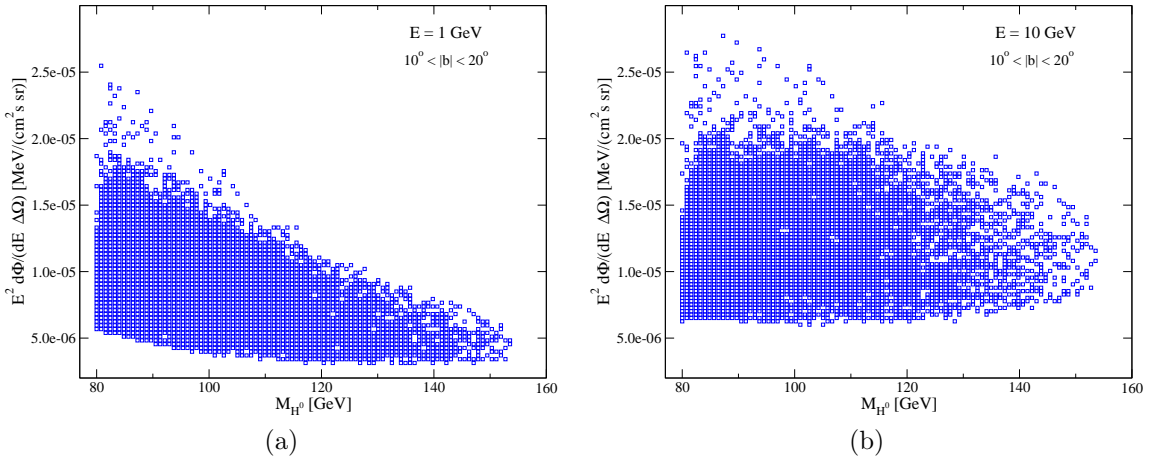

\begin{center} 
\begin{tabular}{cc}
\includegraphics[scale=0.27]{mh0flux1.eps} & \includegraphics[scale=0.27]{mh0flux10.eps}\\
(a) & (b)
\end{tabular}
\caption{The expected gamma ray flux at $1~\gev$ (left) and $10~\gev$ (right) as a function of $\mH$ for our set of viable models. The observation region is $10^\circ<|b|<20^\circ$ in both panels.  \label{fig:gamma}}
\end{center}
\end{figure}

The gamma ray flux from dark matter annihilations originating in a region $\Delta\Omega$ of the sky can be written as
\begin{equation}
\frac{d\Phi_{\mathrm{halo}}}{dE}=\frac12\frac{\sigma \mathrm{v}}{4\pi}r_\odot\frac{\rho_\odot^2}{m_{\mathrm{DM}^2}}\bar J\Delta\Omega\frac{dN}{dE},
\end{equation}
where $r_\odot=8.5\mathrm{~kpc}$ is the distance from the Sun to the
Galactic Center, $\rho_{\odot}=0.39~\gev/\mathrm{cm^3}$ is the local
density of dark matter~\cite{Catena:2009mf}, $dN/dE$ is the photon spectrum, and $\bar J\Delta\Omega$ is given by
\begin{equation}
\bar J\Delta\Omega=\int_{\Delta\Omega}d\Omega(b,l)\int_{\mathrm{los}}\frac{ds}{r_\odot}\left(\frac{\rho_{\mathrm{halo}(r(s,\psi))}}{\rho_\odot}\right)^2\,.
\end{equation}  
For the galactic distribution of dark matter we consider a NFW profile~\cite{Navarro:1995iw},
\begin{equation}
\rho_{\mathrm{NFW}}(r)=\rho_s\frac{r_s}{r}\left(1+\frac{r}{r_s}\right)^{-2}\,,
\end{equation}
where $r_s=20\,$kpc and $\rho_s=0.34\,\gev/\mathrm{cm^3}$. For
definiteness we consider the observation region
$10^\circ<|b|<20^\circ$, for which there is available data from FERMI
-- see~\cite{Abdo:2010nz} and the supplementary online material.
The photon spectrum from dark matter annihilations, $dN/dE$, depends
on the annihilation branching ratios  and has been obtained using micrOMEGAs.

Figure \ref{fig:gamma} shows the differential gamma ray flux at $1~\gev$ (left panel) and at $10~\gev$ (right panel) for our sample of models. For $E_\gamma=10~\gev$, it lies approximately between $5.0\times 10^{-6}$ and $2.5\times 10^{-5}~\mathrm{MeV/(cm^2\,s\, sr)}$, and at slightly smaller values (particularly at large masses) for $1~\gev$. Unfortunately, these fluxes are much smaller than those measured by FERMI~\cite{Abdo:2010nz}, which are consistent with the predictions of cosmic ray models. It is thus not possible to obtain any constraints from this data.

Let us stress that this somewhat disappointing result is consistent
with previous analysis of FERMI data, see e.g.~\cite{FermiDM}.
In them, it had been found that present data from FERMI can be used to
rule out dark matter annihilation cross sections exceeding  $10^{-23}$
or $10^{-24}~\mathrm{cm^3/s}$ (depending on the dark matter mass and
annihilation final state). Such cross sections are much larger than
those expected in typical models of dark matter and, in particular, than
those we have found along the new viable region of the inert doublet model.

\subsection{Direct detection}
\begin{figure}[t]
\begin{center} 
\includegraphics[scale=0.35]{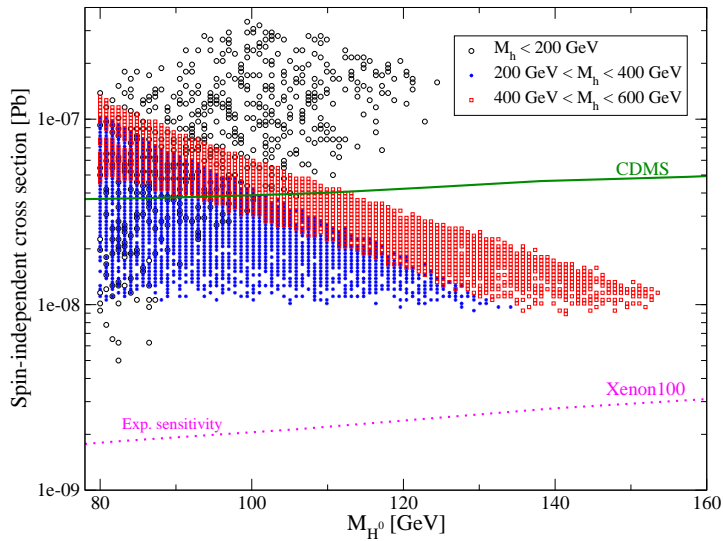}
\caption{The spin-independent direct detection cross section as a function of $\mH$ along the new viable region of the inert doublet model. Three possible ranges of $\mh$ are distinguished by different symbols (and colors). The solid line shows the current bound from CDMS whereas the dotted line corresponds to the expected sensitivity of Xenon100.\label{fig:mh0sdd}}
\end{center}
\end{figure}

Dark matter can also be detected via elastic scattering with terrestrial detectors, the so-called direct detection method. From a particle physics point of view, the quantity that determines the direct detection rate of a dark matter particle is the dark matter-nucleon scattering cross section. In the inert Higgs model, the $H^0N$  scattering process relevant for  direct detection is Higgs-mediated, with a
cross  section, $\sigma_{H^0N}$, given by 
\begin{equation}
\sigma_{H^0N}=\frac{m_r^2}{4\pi}\left( \frac{\lambda_L}{\mH m_h^2}\right)^2f^2m_N^2\,
\end{equation}
where $f\sim 0.3$ is a form factor and $m_r$ is the reduced mass of the system.
Thus, it is proportional to $\lh^2$. Given the rather large values of $\lh$ we found, see figure~\ref{fig:mh0lamh0}, we foresee that  $\sigma_{H^0N}$ will be  significant over the entire new viable region.
Figure \ref{fig:mh0sdd} shows the spin-independent dark matter-nucleon
cross section, $\sigma_{H^0N}$, as a function of $\mH$ for our sample
of models.  $\sigma_{H^0N}$ lies in a narrow range between $3\times
10^{-7}$ and $10^{-8}$ Pb. To ease the  comparison with present
experimental data, the bound from the CDMS
experiment~\cite{Ahmed:2009zw} is also shown (as a solid line) as
well as the expected sensitivity of Xenon100 (dotted line) \cite{Aprile:2010zz}. Points lying above the CDMS line are not consistent with present bounds whereas points above the Xenon100 line are within the reach of that experiment. As observed in the figure, a non-negligible fraction of the new viable region is ruled out by direct detection constraints. In particular, most models  with $\mh<200~\gev$ (though not all of them) have a direct detection cross section exceeding the CDMS bound. No range of $\mH$ is excluded, however. The good news is that all these viable models feature cross sections close to current bounds. They are all well within the sensitivity  of Xenon100. Clearly, direct detection is a promising way of probing this new viable region of the  inert doublet model.

\section{Conclusions}
The inert doublet model is a simple and appealing extension of the
Standard Model that can account for the dark matter. We have
demonstrated, in this paper, the existence of a new, previously
overlooked, viable region of this model featuring dark matter masses
between $M_W$ and about $160~\gev$. In this region of the parameter
space, the correct relic density can be obtained thanks to
cancellations between the different diagrams that contribute to dark
matter annihilation into gauge bosons. In the first part of the paper,
we illustrated how these cancellations come about and analyzed the
dependence of the annihilation cross section and the relic density on
the different parameters of the model. Several examples of viable
models were found both in the light Higgs regime, $\mh<200~\gev$, and
in the heavy Higgs regime, $\mh>200~\gev$. In the second part of the
paper, we made use of Markov Chain Monte Carlo techniques to obtain a
large sample of models within the new viable region. These models were
then analyzed in detail by projecting them onto several
two-dimensional planes. From these projections we learned, for
instance, that for $\mh<200~\gev$ the dark matter particle mass can be
as large as $130~\gev$  whereas  $\lh$ is small and negative
($0>\lh>-0.5$). Dark matter masses above $130~\gev$ (up to $160~\gev$)
can also be found but they feature heavy Higgs masses and positive
values of $\lh$ ($3>\lh>0$). It was also noticed that the allowed
values of $\mC$ and $\mA$ increase with  $\mH$ along the new viable
region, reaching $\mA,\mC\gtrsim 500~\gev$ for $\mH\gtrsim
140~\gev$. In addition, we addressed the issue of the fine-tuning
required to obtain the right relic density and showed that all these
models feature a small or moderate value of the fine-tuning
parameter. Finally, we considered the detection prospects of inert
Higgs dark matter within the new viable region. For collider searches,
the Higgs decay decay rate begins to differ from the Standard Model
one for Higgs masses larger than 200 GeV. The extra contribution
resulting from Higgs decays into inert scalars for $M_h>300$ GeV might
be observed directly at  LHC.
 Regarding indirect
detection, the present annihilation rate, the dominant branching
fractions, and the expected gamma ray flux from dark matter
annihilation were computed. But no bounds could be derived from
current gamma ray data. The direct detection cross section was also
calculated and found to be rather large.  A fraction of the new viable
parameter space, in fact, is already ruled out by present constraints
from CDMS, and the remaining part will be entirely within the expected
sensitivity of Xenon100. Direct detection seems to be, therefore, the
most promising way of probing this new viable region of the inert
doublet model.

\section*{Acknowledgments} 
It is a pleasure to thank M. Gustafsson, E. Nardi and M. Tytgat for comments and suggestions. L. L. H is supported in part by the IISN and by Belgian
Science Policy (IAP VI/11).   
C. E. Y. is supported by the \emph{Juan de la
  Cierva} program of the MICINN of Spain. He acknowledges additional
support from the MICINN Consolider-Ingenio 2010 Programme under grant
MULTIDARK CSD2009-00064, from the MCIINN under Proyecto Nacional
FPA2009-08958, and from the CAM under grant HEPHACOS S2009/ESP-1473. 

\end{document}